\documentclass[preprint,12pt]{elsarticle}

\usepackage{fancyhdr}
\usepackage{fullpage}
\usepackage{amsmath}
\usepackage{amsbsy}
\usepackage{amssymb}
\usepackage{amscd}
\usepackage{amsfonts}
\usepackage{supertabular}
\usepackage{graphics}
\usepackage{verbatim}
\usepackage{epsfig}
\usepackage{xspace}
\usepackage{euscript}
\usepackage{alltt}
\usepackage{boxedminipage}
\usepackage{float}
\usepackage[colorlinks]{hyperref}
\usepackage{color}
\usepackage[all]{xy}
\usepackage{t1enc}
\usepackage{times,exscale}
\usepackage{graphicx,calc}
\usepackage{subfig}
\usepackage[ruled,vlined]{algorithm2e}
\usepackage{epstopdf}
\usepackage{array,multirow}

\def\bR{{\mathbf{R}}}

\def\bx{{\mathbf{x}}}

\def\fd{{g(\mathbf{H},\mu,\sigma)}}
\def\fdp{{g(\mathbf{H}_p,\mu,\sigma)}}
\def\fdps{{g(\hat{\mathbf{H}}_p,\hat{\mu}_p,\hat{\sigma}_p)}}

\journal{arXiv}

\begin{document}

\begin{frontmatter}

\title{On nearsightedness in metallic systems for $\mathcal{O}(N)$ Density Functional Theory calculations: A case study on Aluminum}
\author[gatech]{Phanish Suryanarayana\corref{cor}}
\address[gatech]{College of Engineering, Georgia Institute of Technology, Atlanta, GA 30332, USA}
\cortext[cor]{Corresponding Author (\it phanish.suryanarayana@ce.gatech.edu) }

\begin{abstract}
We investigate the locality of electronic interactions in aluminum as a function of smearing/electronic temperature in the context of $\mathcal{O}(N)$ Density Functional Theory calculations. Specifically, we determine the convergence in energy and atomic forces with truncation region size for smearing of $0.001-0.15$ Ha. We find exponential convergence accompanied by a rate that increases sub-linearly with smearing, with truncation region sizes of $48-64$ Bohr required to achieve chemical accuracy for typical smearing values of $0.001-0.01$ Ha. This translates to $\mathcal{O}(N)$ scaling for systems larger than $\mathcal{O}(1000)$ atoms.
\end{abstract}


\end{frontmatter}

\section{Introduction}
Density Functional Theory (DFT) \cite{Hohenberg,Kohn1965} is widely used ab-initio method (no empirical or adjustable parameters) for understanding and predicting a diverse range of materials properties. The main computational bottleneck in DFT simulations is the calculation of the eigenfunctions and eigenvalues corresponding to the occupied spectrum of the Kohn-Sham eigenproblem. Since these eigenfunctions---number proportional to the number of atoms $N$---need to be orthogonal, the overall computational complexity of DFT calculations is $\mathcal{O}(N^3)$ \cite{Martin2004,NumAnalysis2003}, which severely restricts the size of systems that can be studied. 

In order to overcome the $\mathcal{O}(N^3)$ scaling bottleneck, a number of real-space solution strategies have been developed over the past two decades that scale as $\mathcal{O}(N)$ (see, e.g., \cite{Goedecker,Bowler2012} and references therein), with mature implementations of the key ideas now available \cite{SIESTA,Conquestref,ONETEP,FEMTECKref,MGMolref,BigDFTref,OpenMXweb,FreeONref}. These approaches circumvent the calculation of the Kohn-Sham orbitals and directly determine the quantities of interest like electron density, energy, and atomic forces. However, nearly all of the developed $\mathcal{O}(N)$ techniques assume the presence of a bandgap in the electronic structure, which makes them unsuitable for the study of metallic systems \cite{Goedecker,Bowler2012} as well as systems that undergo transition between insulating and metallic \cite{kerszberg2015ab}.  

A key property exploited by $\mathcal{O}(N)$ methods is the \emph{nearsightedness principle} \cite{prodan2005nearsightedness}, i.e., the locality of electronic interactions in real-space. This manifests itself in the exponential decay of the real-space density matrix for insulators as well as metallic systems at finite electronic temperature/occupation-smearing \cite{goedecker1998decay,ismail1999locality,zhang2001properties,taraskin2002spatial,benzi2013decay}. The truncation that can be employed in practice, while maintaining the desired accuracy, determines the prefactor of $\mathcal{O}(N)$ methods as well as the system sizes at which linear scaling can be achieved in practical calculations. While this has been studied previously for insulators \cite{bowler2002recent,skylaris2007achieving}, the dependence of truncation region size on smearing for a given accuracy in metallic systems has not been carefully studied heretofore \cite{aarons2016perspective}. 

In this work, we employ the recently developed $\mathcal{O}(N)$ Spectral Quadrature (SQ) method \cite{suryanarayana2013spectral,pratapa2016spectral}---identically applicable to both insulating and metallic systems---to study the locality of electronic interactions in aluminum (a prototypical metallic system) for various smearings. We find that the convergence of energy and atomic forces with truncation region size is exponential, with an associated rate whose growth increases with smearing, while remaining sub-linear. Notably, truncation region sizes of $48-64$ Bohr are required to achieve chemical accuracy for typical smearing values of $0.001-0.01$ Ha. We also find through comparison with lithium and molybdenum that the difference in truncation region sizes for various metallic systems is likely to be consequence of the difference in the prefactor rather than the convergence rate. 

The remainder of this paper is organized as follows. We outline $\mathcal{O}(N)$ DFT in Section \ref{Sec:ONDFT} and the Clenshaw-Curtis SQ method in Section \ref{Sec:CCSQDFT}. Next, we study the nearsightedness of aluminum in the context of practical $\mathcal{O}(N)$ DFT calculations in Section \ref{Sec:Results}. Finally, we provide concluding remarks in Section \ref{Sec:Conclusions}. 

\section{$\mathcal{O}(N)$ Density Functional Theory} \label{Sec:ONDFT}
In Density Functional Theory (DFT), the Kohn-Sham eigenproblem can be rewritten as \cite{anantharaman2009existence,pratapa2016spectral}
\begin{equation} \label{Eqn:SelfCons:DensityMatrix}
\mathcal{D}  = g(\mathcal{H},\mu,\sigma) = \left( 1 + \exp \left(\frac{\mathcal{H}-\mu \mathcal{I}}{\sigma} \right)   \right)^{-1} \,,
\end{equation} 
where $\mathcal{D}$ is the density matrix, $g$ is the Fermi-Dirac function, $\mu$ is the Fermi level, $\sigma$ is value of the smearing, and $\mathcal{H}$ is the Hamiltonian:
\begin{equation} \label{Eqn:Hamiltonian}
\mathcal{H} = -\frac{1}{2} \nabla^2 + V_{xc} + \phi + \mathcal{V}_{nl} \,.
\end{equation}
Above, $V_{xc}$ is the exchange-correlation potential, $\phi$ is the electrostatic potential, and $\mathcal{V}_{nl}$ is the nonlocal pseudopotential. The electrostatic potential $\phi$ is the solution of the Poisson equation \cite{Pask2005,Phanish2010,Phanish2011}
\begin{equation} \label{Eqn:PoissonEqn}
-\frac{1}{4 \pi} \nabla^2 \phi(\bx,\bR) = \rho_{\mathcal{D}}(\bx) + b(\bx,\bR) \,,
\end{equation}
where $\bR = \{\bR_1, \bR_2, \ldots, \bR_N \}$ denotes the position of the nuclei, $\rho_{\mathcal{D}}(\bx) = 2 \mathcal{D}(\bx,\bx)$ is the electron density, and $b$ is the total pseudocharge density of the nuclei. The Fermi level $\mu$ is determined by solving for the constraint on the total number of electrons, i.e., $2 \mathrm{Tr} (\mathcal{D}) = N_e$, where $\mathrm{Tr}$ denotes the trace.  

Once the electronic ground-state has been determined, the free energy takes the form \cite{pratapa2016spectral}
\begin{eqnarray} 
\mathcal{F} (\bR) & = & 2\text{Tr}(\mathcal{D} \mathcal{H})  + E_{xc}(\rho_{\mathcal{D}}) - \int V_{xc}(\rho_{\mathcal{D}}(\bx)) \rho_{\mathcal{D}}(\bx) \, \mathrm{d\bx}  \nonumber \\
& &  + \frac{1}{2} \int (b(\bx,\bR)-\rho_{\mathcal{D}}(\bx)) \phi(\bx,\bR) \, \mathrm{d\bx} - E_{self}(\bR)  + E_{corr}(\bR) \label{DM:groundstate} \\
& & + 2 \sigma \text{Tr}\left( \mathcal{D} \log \mathcal{D} + (\mathcal{I}-\mathcal{D}) \log (\mathcal{I}-\mathcal{D}) \right) \,, \nonumber
\end{eqnarray}
where $E_{xc}$ is the exchange-correlation energy, $E_{self}$ is the self energy associated with the pseudocharges, and $E_{corr}$ is the electrostatic correction for overlapping pseudocharges. Thereafter, the force on the $I^{th}$ nucleus can be written as \cite{pratapa2016spectral}
\begin{equation} \label{Eqn:forces} 
\mathbf{f}_I = \sum_{I'} \int \nabla b_{I'}(\bx,\bR_{I'})(\phi(\bx,\bR)-V_{I'}(\bx,\bR_{I'})) \, \mathrm{d\bx} + {\mathbf{f}}_{I,corr} - 4\text{Tr}\left(\nabla \mathcal{D} \, \mathcal{V}_{nl,I}\right)  \,,
\end{equation}
where the summation index $I'$ runs over the $I^{th}$ atom and its periodic images, $b_{I}$ is the pseudocharge density of the $I^{th}$ nucleus that generates the potential $V_{I}$, $\mathbf{f}_{I,corr}$ is the electrostatic force correction arising from overlapping pseudocharges, and $\mathcal{V}_{nl,I}$ is the nonlocal pseudopotential associated with the atom.  

The real-space density matrix has exponential decay for insulators as well as metallic systems at finite temperature \cite{goedecker1998decay,benzi2013decay}. Linear-scaling methods exploit this property for the $\mathcal{O}(N)$ calculation of the ground-state electron density, energy and atomic forces given in Eqns.~\ref{Eqn:SelfCons:DensityMatrix}, \ref{DM:groundstate}, and \ref{Eqn:forces}, respectively. Note that in the above description for DFT, we have employed a local reformulation of the electrostatics \cite{Phanish2012,ghosh2016higher} to enable $\mathcal{O}(N)$ scaling for the complete DFT problem. 

\section{Clenshaw-Curtis Spectral Quadrature method} \label{Sec:CCSQDFT}
The Clenshaw-Curtis Spectral Quadrature (SQ) method is an $\mathcal{O}(N)$  technique for performing Density Functional Theory (DFT) calculations that is identically applicable to both insulating and metallic systems \cite{suryanarayana2013spectral,pratapa2016spectral}. It has been formulated in terms of the finite-difference discretization in order to exploit the locality of electronic interactions in real space, enable systematic convergence, and facilitate large-scale parallel implementation. In the SQ method, the quantities of interest are expressed as bilinear forms or sums over bilinear forms, which are then approximated by spatially localized Clenshaw-Curtis quadrature rules. In doing so, the two additional parameters introduced into the calculation are the order of quadrature ($n_{pl}$) and the size of the truncation region ($2 R_{cut}$).

As an example, in each iteration of the self-consistent field (SCF) method, the electron density at the $p^{th}$ finite-difference node is written as
\begin{eqnarray} 
\rho_{p} & = & \frac{2}{h^3} \mathbf{v}_p^T \fd \mathbf{v}_p \approx \frac{2}{h^3} \mathbf{w}_p^T \fdp \mathbf{w}_p = \frac{2}{h^3} \mathbf{w}_p^T \fdps \mathbf{w}_p  \nonumber \\
& \approx & \frac{2}{h^3} \mathbf{w}_p^T \left(\sum_{j=0}^{n_{pl}} c_p^j(\mu) T_j(\hat{\mathbf{H}}_p) \right) \mathbf{w}_p = \frac{2}{h^3} \sum_{j=0}^{n_{pl}} c_p^j(\mu) \rho_p^j \,,
\end{eqnarray}
where $h$ is the mesh-size; $\mathbf{v}_p$ is a column vector with the only non-zero entry being $1$ in the $p^{th}$ position; $\mathbf{w}_p$ is the restriction of $\mathbf{v}_p$ to the truncation region of the $p^{th}$ node; $\hat{\mathbf{H}}_p$ is the nodal Hamiltonian---restriction of the Hamiltonian to the truncation region of the  $p^{th}$ node---which is scaled and shifted such that its spectrum lies in $[-1,1]$; and $c_p^j$ are the Chebyshev expasion coefficients for the function $g$, with the Fermi level $\mu$ chosen to satisfy the constraint on the total number of electrons, i.e.,
\begin{equation}
2 \sum_{p} \sum_{j=0}^{n_{pl}} c_p^j(\mu) \rho_p^j = N_e \,.
\end{equation} 
Similarly, the band structure energy, electronic entropy, and nonlocal component of the atomic force can be expressed in terms of nodal quantities, details of which can be found in previous work \cite{pratapa2016spectral}. When combined with a local reformulation of the electrostatics \cite{Pask2005,Pask2012,Phanish2012,ghosh2016higher}, the SQ method enables the $\mathcal{O}(N)$ evaluation of the electron density, energy, and atomic forces for insulating as well as metallic systems.

In this work, we employ the \emph{infinite-cell} version of the SQ method, wherein the results corresponding to the infinite crystal are obtained without recourse to Brillouin zone integration or large supercells \cite{pratapa2016spectral}. Specifically, rather than employ Bloch boundary conditions on the orbitals in the unit cell, zero-Dirichlet (or equivalently periodic) boundary conditions are prescribed at infinity, and the relevant components of the density matrix associated with any spatial point within the unit cell are calculated by utilizing the potential within the truncation region surrounding that point. This is theoretically equivalent to the calculation of the density matrix in all of space while employing truncation in a region of size $2R_{cut}$ throughout the $\mathcal{O}(N)$ method, and then utilizing the components of the density matrix corresponding to the spatial points within the unit cell for the calculation of the electron density, energy, and atomic forces. 

\section{Results and discussion} \label{Sec:Results}
In this section, we utilize the Spectral Quadrature (SQ) method to study the nearsightedness of matter in the context of practical $\mathcal{O}(N)$ Density Functional Theory (DFT) calculations. We focus on aluminum in this work, since it is a prototypical metallic system. Within DFT, we utilize the Perdew-Wang parametrization \cite{perdew1992accurate} of the correlation energy calculated by Ceperley-Alder \cite{Ceperley1980} and norm-conserving Troullier-Martins pseudopotentials \cite{Troullier}. We accelerate the convergence of the self-consistent field (SCF) method using the Periodic Pulay mixing scheme \cite{banerjee2016periodic} with restarts \cite{pratapa2015restarted}. In each SCF iteration, we solve the Poisson equation for the electrostatic potential using the Alternating Anderson-Jacobi (AAJ) method \cite{pratapa2016anderson,suryanarayana2016alternating} and determine the Fermi level using Brent's method. 

In all the simulations, we utilize a twelfth-order finite-difference discretization with mesh-size $h$ chosen such that the energy and atomic forces are converged---with respect to highly converged plane-wave results computed by ABINIT \cite{ABINIT} as well as real-space results computed using SPARC \cite{ghosh2016sparc1,ghosh2016sparc2}---to within $10^{-4}$ Ha/atom and $10^{-4}$ Ha/Bohr, respectively.\footnote{It has been verified that the results presented in this work are converged with respect to mesh-size.}  We choose the quadrature order within the SQ method so that the resulting errors in the energy and atomic forces are within $10^{-6}$ Ha/atom and $10^{-6}$ Ha/Bohr, respectively. Here and below, the error in the energy and atomic forces denotes the magnitude of the difference and the maximum difference in any component, respectively. In addition, all coordinates are defined with respect to the origin located at the corner of the unit cell. 

\subsection{Nearsightedness in aluminum} \label{Subsec:Al}
We consider a $4$-atom face-centered cubic (FCC) unit cell of aluminum at the equilibrium lattice constant of $7.78$ Bohr, with the atom located at [$3.89$ $3.89$ $0.00$] Bohr moved to [$3.74$ $3.49$ $0.37$] Bohr.\footnote{The nature of the results presented in this work remain unchanged even when a perfect FCC aluminum crystal is considered.} In addition, we choose the following values for the smearing: $\sigma = 0.001$, $0.005$, $0.010$, $0.050$, $0.100$, and $0.150$ Ha. These values encompass the wide range of temperatures encountered in Density Functional Theory (DFT) calculations, ranging from ambient to warm dense matter \cite{kresse1993ab,renaudin2003aluminum}. 

\begin{figure}[h!]
\centering
\subfloat[Energy]{\label{Fig:EnergyConvergence_Al}\includegraphics[keepaspectratio=true,width=0.49\textwidth]{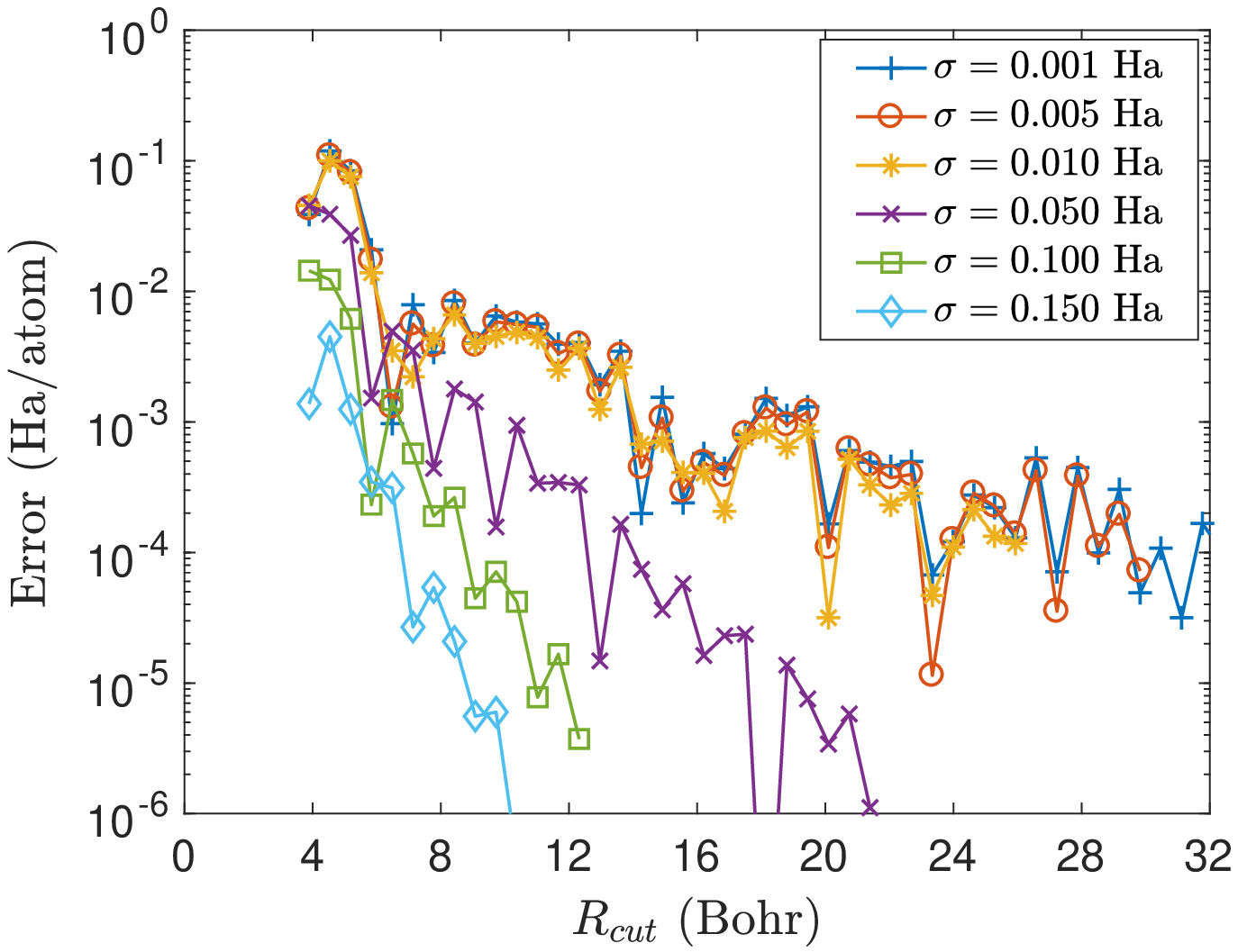}}
\subfloat[Atomic forces]{\label{Fig:ForceConvergence_Al}\includegraphics[keepaspectratio=true,width=0.49\textwidth]{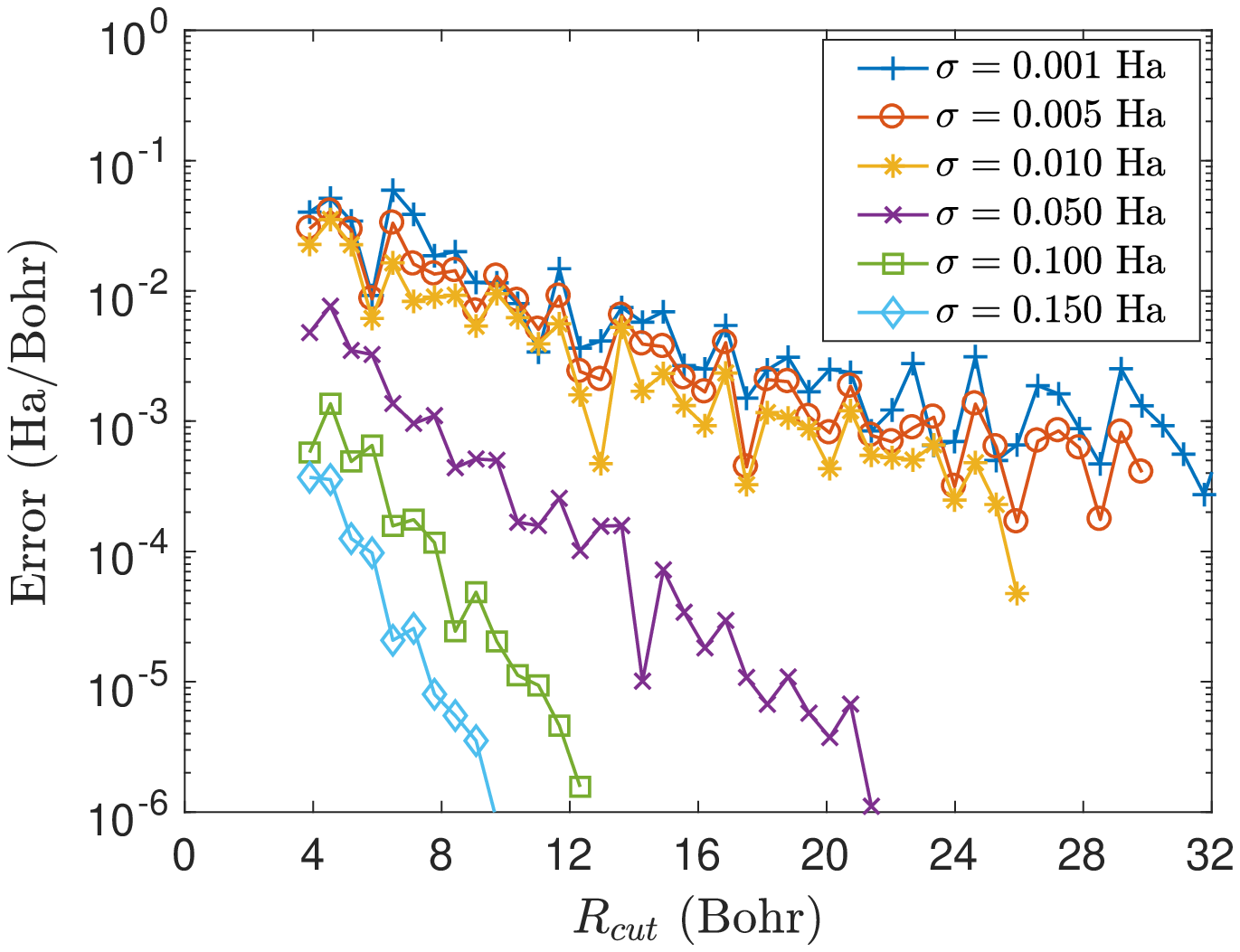}}
\caption{Convergence of energy and atomic forces in aluminum with respect to truncation region size ($2 R_{cut}$) for various values of the smearing ($\sigma$). The results obtained by SPARC \cite{ghosh2016sparc1,ghosh2016sparc2} via diagonalization for the same mesh-size and finite-difference order are used as reference.}
\label{Fig:Error:Smearing}
\end{figure}

\begin{figure}[h!]
\centering
\subfloat[Convergence rate]{\label{Fig:rate_vs_smearing}\includegraphics[keepaspectratio=true,width=0.49\textwidth]{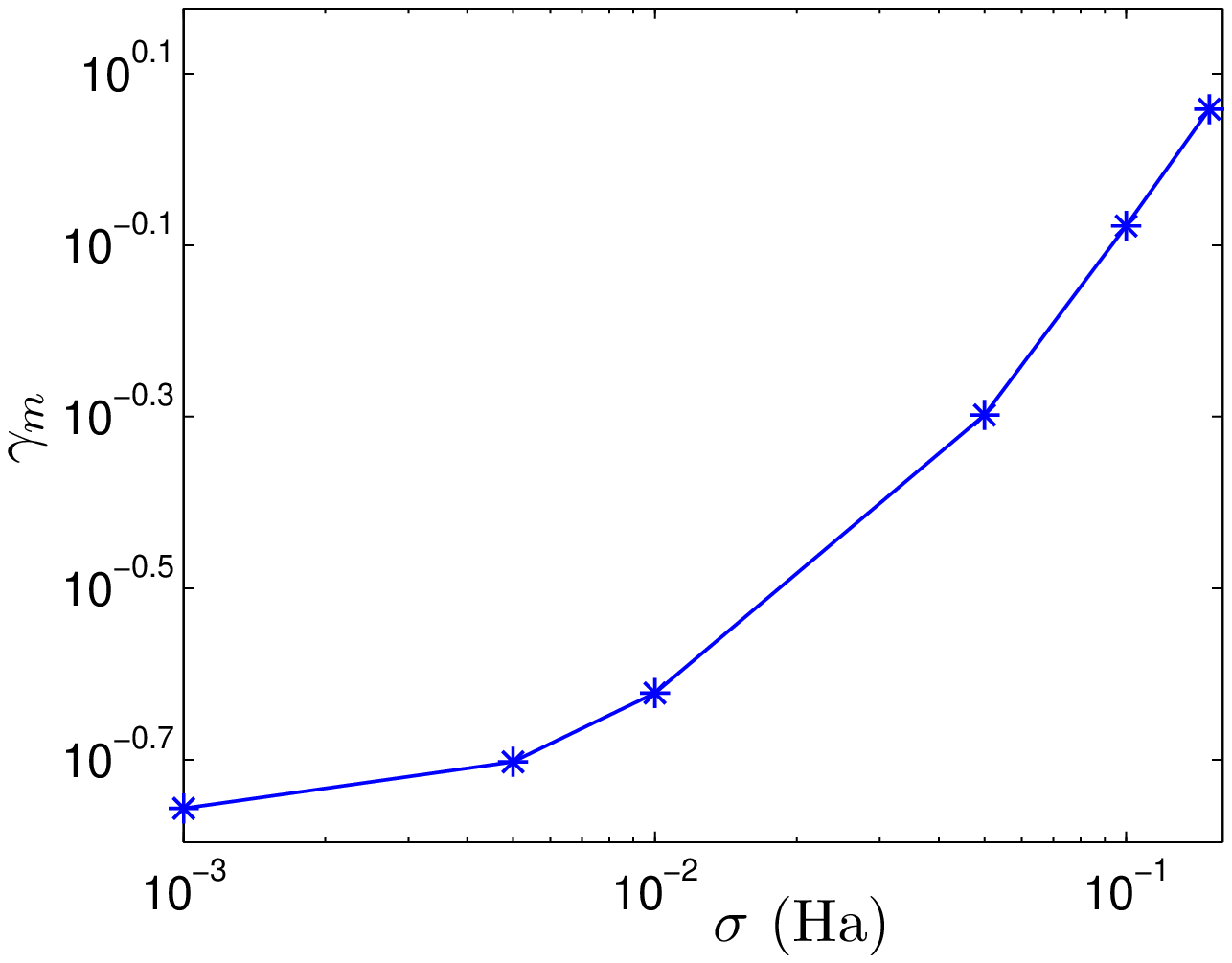}}
\subfloat[Prefactor]{\label{Fig:pre_vs_smearing}\includegraphics[keepaspectratio=true,width=0.49\textwidth]{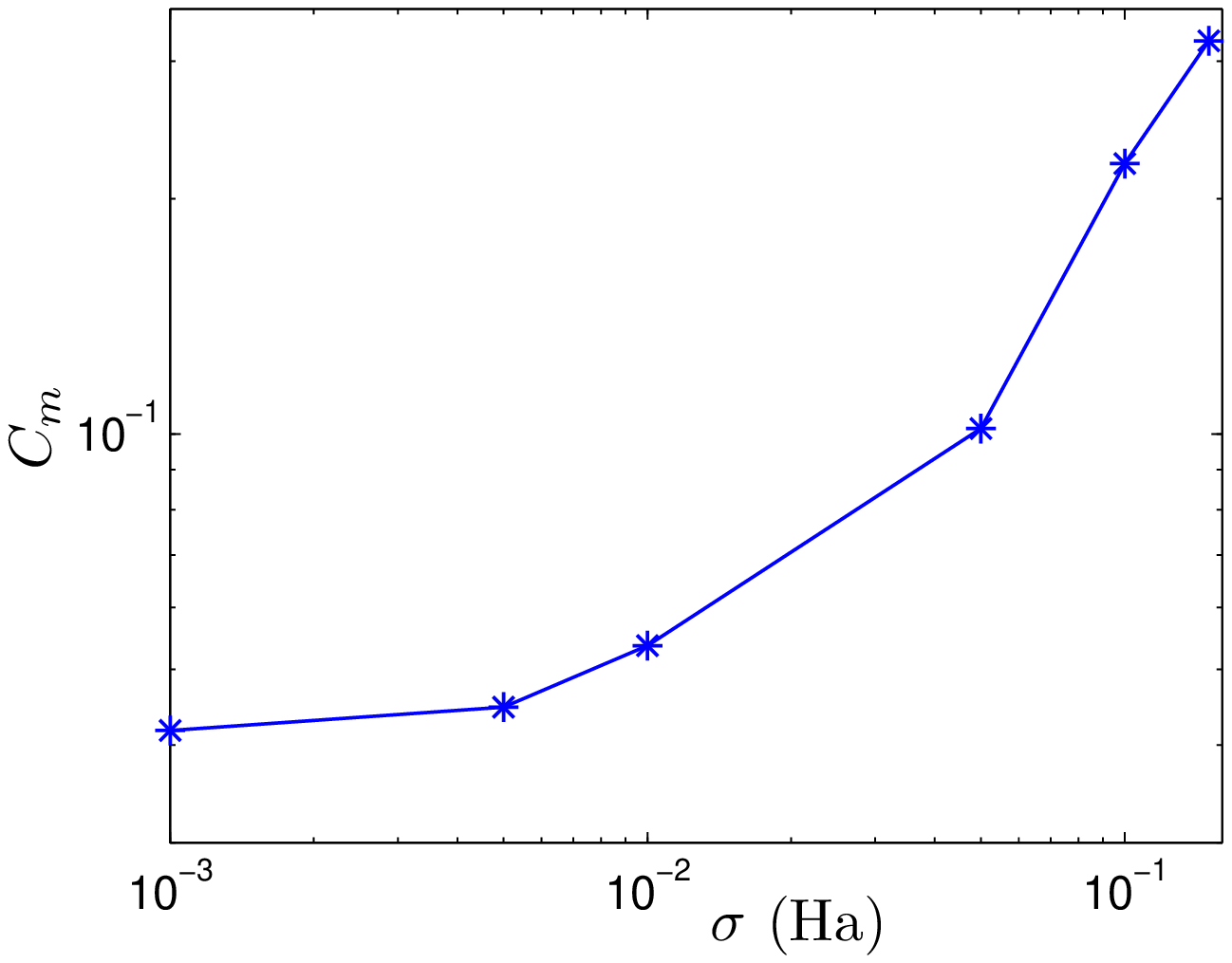}}
\caption{Average (between the energy and the atomic forces) rate of convergence $\gamma_m$ and the associated prefactor $C_m$ as a function of  the smearing $\sigma$ for the aluminum system.}
\label{Fig:Pre:Rate:Smearing}
\end{figure}

In Fig. \ref{Fig:Error:Smearing}, we plot the variation of the error in the energy and atomic forces as a function of $R_{cut}$ for the aforementioned values of smearing. The results obtained by SPARC \cite{ghosh2016sparc1,ghosh2016sparc2} (via diagonalization) for the same mesh-size and finite-difference order are used as reference. We observe that there is exponential convergence in the energy and atomic forces at all smearings, with relatively minor variations between $\sigma=0.001-0.01$ Ha. Next, we fit the above data to the function
\begin{equation}
\text{Error} = C \exp(- \gamma R_{cut}) \,,
\end{equation}
where $\gamma$ denotes the convergence rate, and $C$ is the associated prefactor. In Fig. \ref{Fig:Pre:Rate:Smearing}, we plot the values of $\gamma_m$ and $C_m$---values of $C$ and $\gamma$ averaged between the energy and atomic forces---as a function of $\sigma$. We observe that both $\gamma_m$ and $C_m$ increase monotonically as $\sigma$ is increased, with similar rates ($\gamma_m \propto \sigma^{0.07-0.77}$ and $C_m \propto \sigma^{0.04-0.89}$). Overall, we conclude that the convergence in the energy and atomic forces with truncation region size is exponential, with the rate demonstrating a growth that increases with smearing, while remaining sub-linear. 

\begin{figure}[h!]
\centering
\subfloat[{$\bx_0 = [0.00 \,\, 0.00 \,\, 0.00]$ Bohr}]{\label{Fig:decay:x0}\includegraphics[keepaspectratio=true,width=0.49\textwidth]{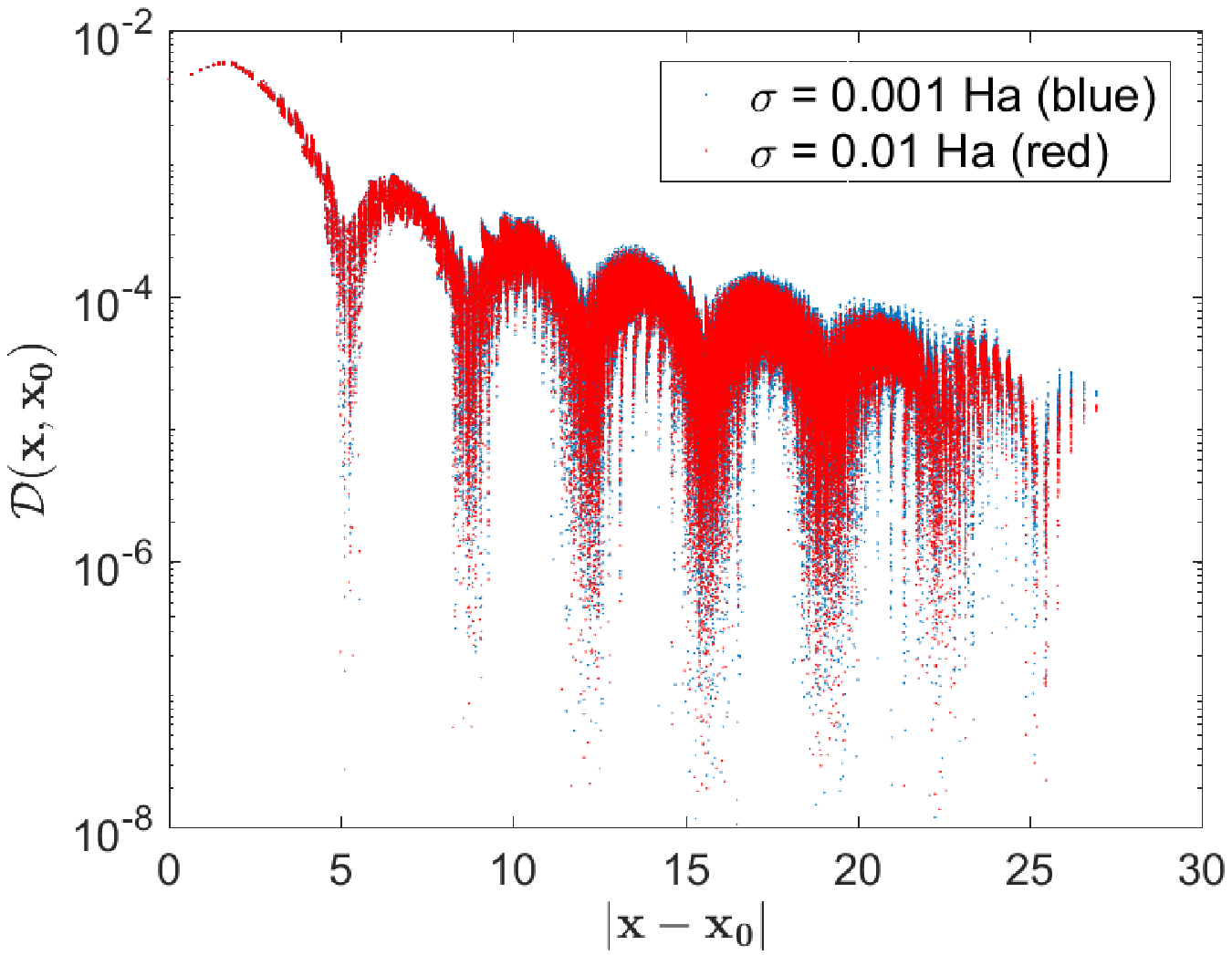}}
\subfloat[{$\bx_1 = [3.89 \,\, 3.89 \,\, 3.89]$ Bohr}]{\label{Fig:decay:x1}\includegraphics[keepaspectratio=true,width=0.49\textwidth]{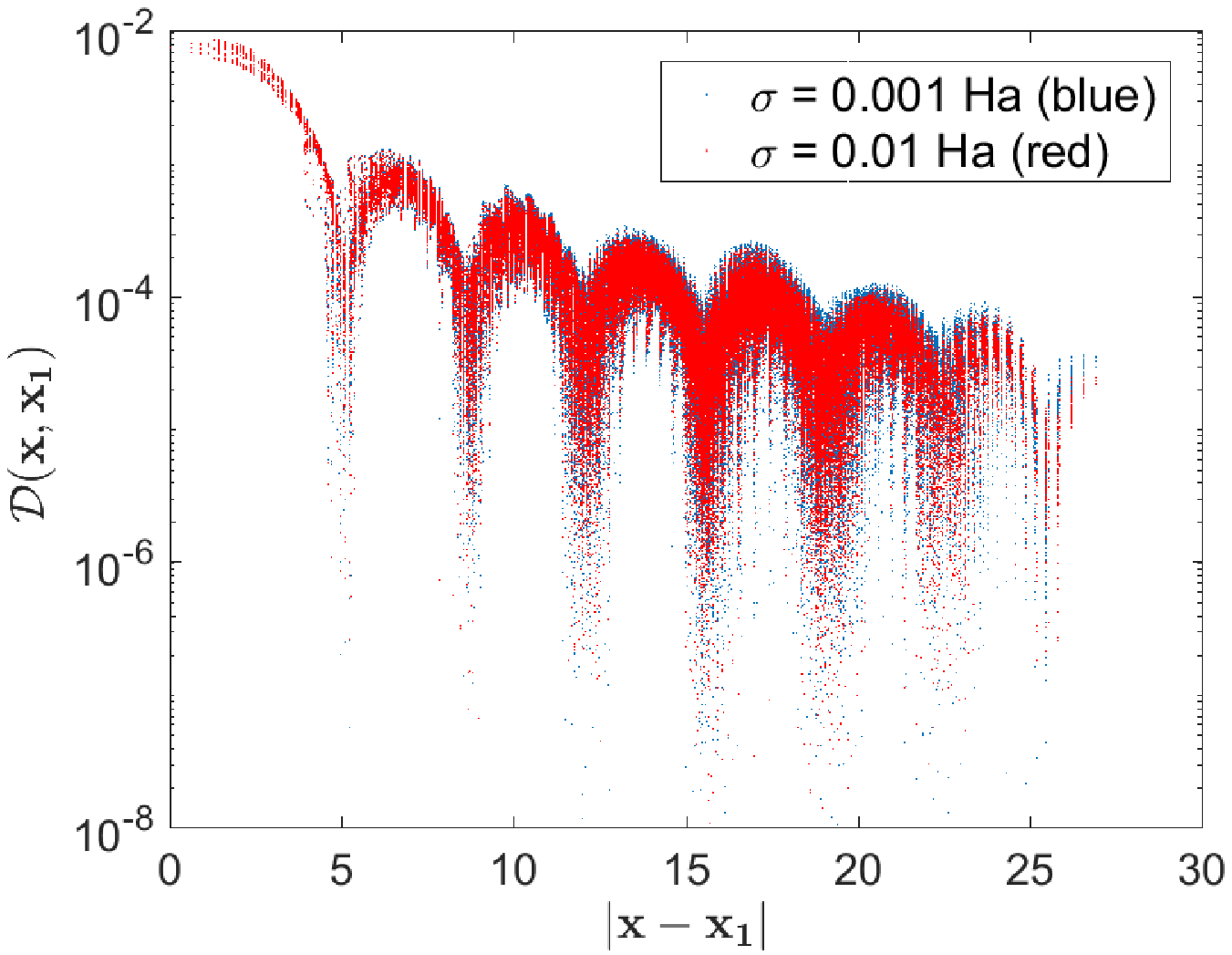}}
\caption{Components of the density matrix for the aluminum system.}
\label{Fig:Decay:DM}
\end{figure}

In previous theoretical work \cite{goedecker1998decay,ismail1999locality,benzi2013decay}, it has been predicted that for finite values of smearing, the density matrix has exponential decay with a rate that increases linearly (an upper bound) with the smearing \cite{benzi2013decay}. Though we obtain exponential convergence of the energy and atomic forces, the mathematical results are not directly applicable in the context of an $\mathcal{O}(N)$ method. This is because of the continuous truncation of the density matrix (explicitly or implicitly) throughout the $\mathcal{O}(N)$ approach, rather than truncation once the complete density matrix is calculated. In order to verify this, we plot in Fig. \ref{Fig:Decay:DM} the components of the density matrix associated with the spatial points $\bx_0 = [0.00 \,\, 0.00 \,\, 0.00]$ Bohr and $\bx_1 = [3.89 \,\, 3.89 \,\, 3.89]$ Bohr for $\sigma=0.001$ Ha and $\sigma=0.01$ Ha.\footnote{We have verified that the decay of $\mathcal{D}(\bx,\bx_0)$ and $\mathcal{D}(\bx,\bx_1)$ are representative of other spatial points within the unit cell.} The decay rates of $\mathcal{D}(\bx,\bx_0)$ at these smearings are $0.171$ and $0.178$, respectively. The corresponding rates for $\mathcal{D}(\bx,\bx_1)$ are $0.171$ and $0.177$, respectively. However, the associated convergence rates of the energy and forces in the $\mathcal{O}(N)$ SQ method are $0.175$ and $0.239$, respectively. These results indicate that the variation in the decay rate of the density matrix with smearing is not in exact correspondence with the convergence of energy and atomic forces in an $\mathcal{O}(N)$ method. 

The smearing is typically chosen to correspond to the physical temperature in the DFT simulation. However, in order to improve the efficiency and stability of the calculations, this constraint can be relaxed. Specifically, a larger smearing can be chosen, while maintaining the desired accuracy in the energy and forces. For example, consider the reference energy and forces at the various smearings for the aluminum system listed in Table \ref{Table:energy:force}. It is clear that the results for $\sigma=0.01$ Ha provide a good approximation to those at $\sigma=0.001$ Ha. Though the truncation region size that needs to be chosen at these two smearings is not noticeably different (Fig. \ref{Fig:Error:Smearing}), close to an order of magnitude speedup can be obtained due to the reduction in the quadrature order ($n_{pl}$) required within the SQ method. 

\begin{table}[h!]
\centering
\begin{tabular}{|c|c|cccc|}
\hline 
$\sigma$ (Ha) & $\mathcal{F}$ (Ha/atom) & $|\mathbf{f}_1|$ (Ha/Bohr) & $|\mathbf{f}_2|$ (Ha/Bohr) & $|\mathbf{f}_3|$ (Ha/Bohr) & $|\mathbf{f}_4|$ (Ha/Bohr) \\ 
\hline \hline
$0.001$ & -2.0859 & 0.0112 & 0.0107 & 0.0156 & 0.0219 \\ 
\hline
$0.005$ & -2.0864 & 0.0111 & 0.0105 & 0.0154 & 0.0220 \\ 
\hline
$0.010$ & -2.0877 & 0.0109 & 0.0103 & 0.0153 & 0.0219 \\ 
\hline
$0.050$ & -2.1313 & 0.0102 & 0.0096 & 0.0147 & 0.0226  \\ 
\hline
$0.100$ & -2.2633 & 0.0116 & 0.0108 & 0.0166 & 0.0274 \\ 
\hline
$0.150$ & -2.4666 & 0.0144 & 0.0134 & 0.0201 & 0.0340 \\ 
\hline
\end{tabular} 
\caption{Reference energy and atomic forces for the aluminum system at various values of smearing ($\sigma$).}
\label{Table:energy:force}
\end{table}

\begin{figure}[h!]
\centering
\subfloat[Energy]{\label{Fig:Comp:erfc:fd:energy}\includegraphics[keepaspectratio=true,width=0.49\textwidth]{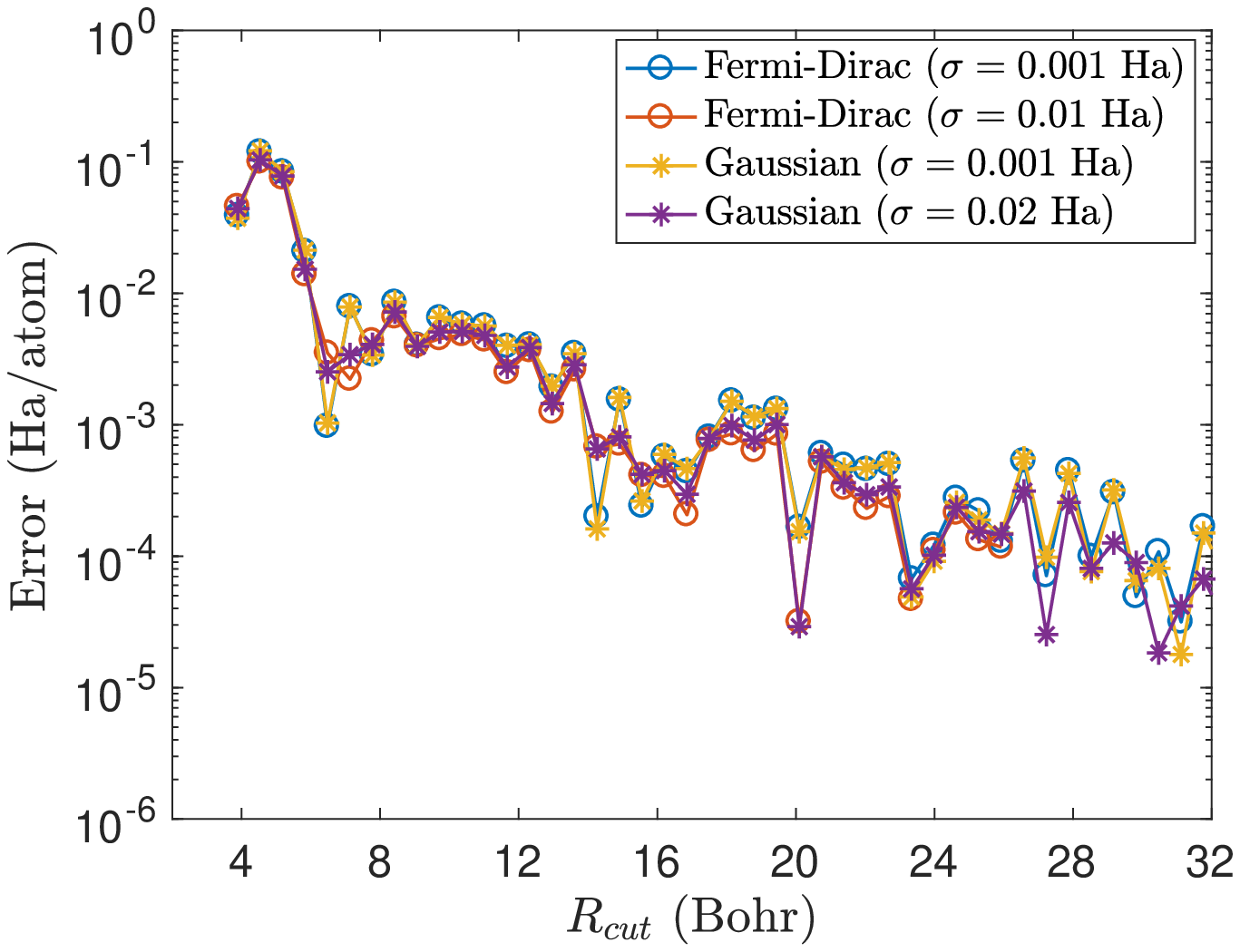}}
\subfloat[Atomic forces]{\label{Fig:Comp:erfc:fd:forces}\includegraphics[keepaspectratio=true,width=0.49\textwidth]{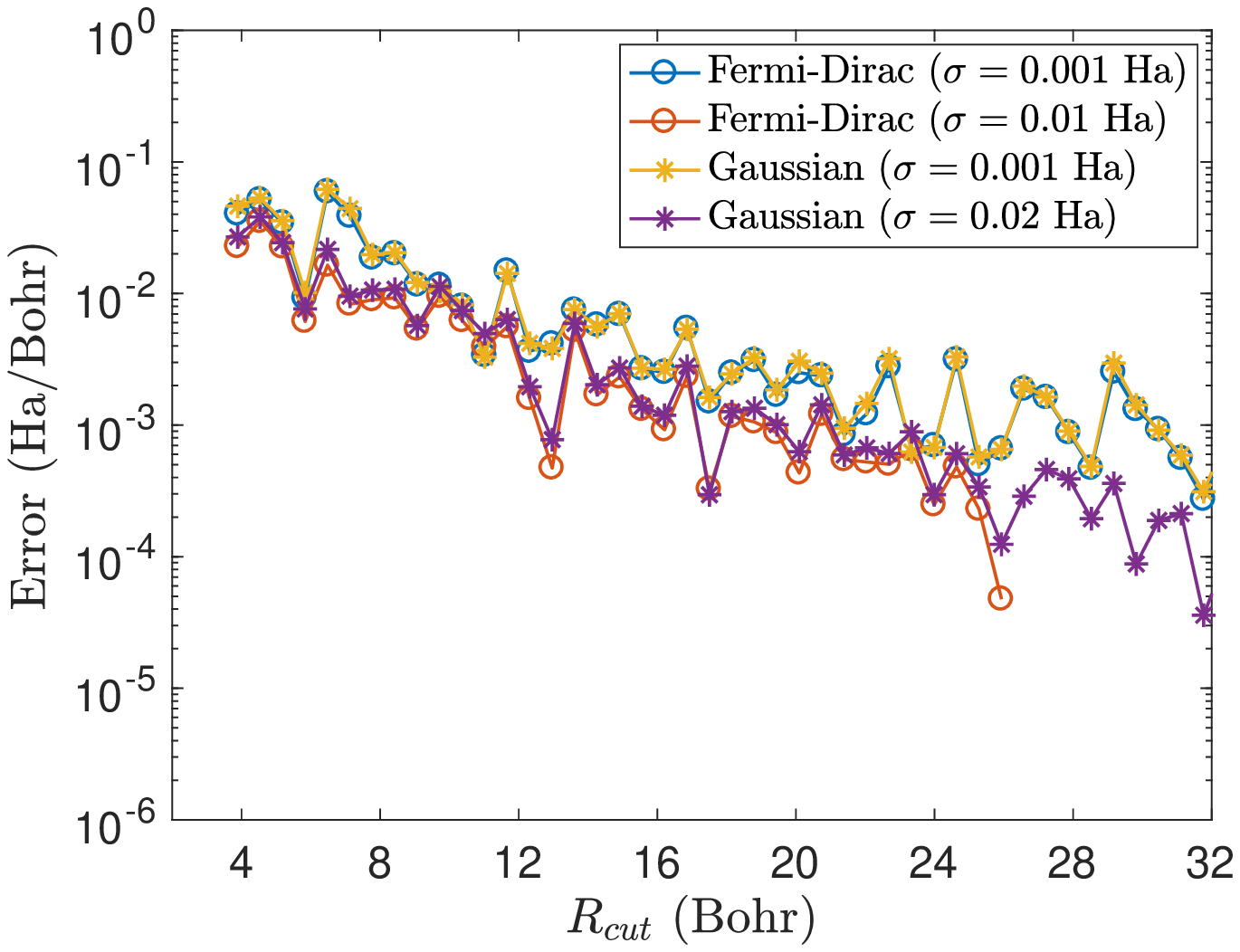}}
\caption{Convergence of energy and atomic forces in aluminum with respect to truncation region size ($2R_{cut}$) for the Fermi-Dirac and Gaussian smearing functions. The results obtained by SPARC \cite{ghosh2016sparc1,ghosh2016sparc2} via diagonalization for the same mesh-size and finite-difference order are used as reference. }
\label{Fig:erfc:fd}
\end{figure}

Another option to relax the constraint on the physical temperature is to employ alternate smearing functions, e.g. Gaussian smearing \cite{VASP}, wherein 
\begin{equation}
g(\mathcal{H},\mu,\sigma)= \frac{1}{2} erfc \left( \frac{\mathcal{H} - \mu \mathcal{I}}{\sigma} \right) \,,  
\end{equation}
$erfc$ being the complementary error function. Since $erfc$ is steeper than the Fermi-Dirac function, larger values of smearing can be employed, e.g. Gaussian smearing with $\sigma=0.02$ Ha provides similar accuracy as Fermi-Dirac smearing with $\sigma=0.01$ Ha, while considering the Fermi-Dirac smearing with $\sigma=0.001$ Ha as reference. In Fig. \ref{Fig:erfc:fd}, we present the convergence of the energy and atomic forces with $R_{cut}$ for the Fermi-Dirac and Gaussian smearing functions. We observe similar convergence for both choices, with no noticeable gain from the ability to choose a larger value of $\sigma$ in Gaussian smearing. However, a lower value of quadrature can be employed in SQ, mainly because $erfc$ is an entire function in the complex plane \cite{suryanarayana2013spectral}. Overall, relaxing of the constraint on the physical temperature does not have any significant influence on the required truncation region size, but can provide improvement in the efficiency of the calculations because of the enhanced smoothness of the smearing function.  
        
\subsection{Comparison of nearsightedness in aluminum with lithium and molybdenum}\label{Subsec:ComparisonLiMo}
We now compare the results obtained for aluminum with lithium and molybdenum for a smearing of $\sigma = 0.01$ Ha and $\sigma = 0.15$ Ha, respectively. We consider a $2$-atom body-centered cubic (BCC) unit cell of lithium at equilibrium lattice constant of $6.24$ Bohr, with the  atom positioned at [$3.12$ $3.12$ $3.12$] Bohr moved to [$3.99$ $2.49$ $3.61$] Bohr; and $2$-atom BCC unit cell of molybdenum at equilibrium lattice constant of $5.97$ Bohr, with the atom located at [$2.985$ $2.985$ $2.985$] Bohr moved to [$2.815$ $3.215$ $3.185$] Bohr. We employ mesh-sizes of $0.52$ and $0.299$ Bohr for the lithium and molybdenum systems, respectively, which results in energy and forces that are with respect to highly accurate plane-wave results to within $10^{-4}$ Ha/atom and $10^{-4}$ Ha/Bohr, respectively.  

\begin{figure}[h!]
\centering
\subfloat[Energy]{\label{Fig:Comparison_energy_Al_Mo_Li}\includegraphics[keepaspectratio=true,width=0.49\textwidth]{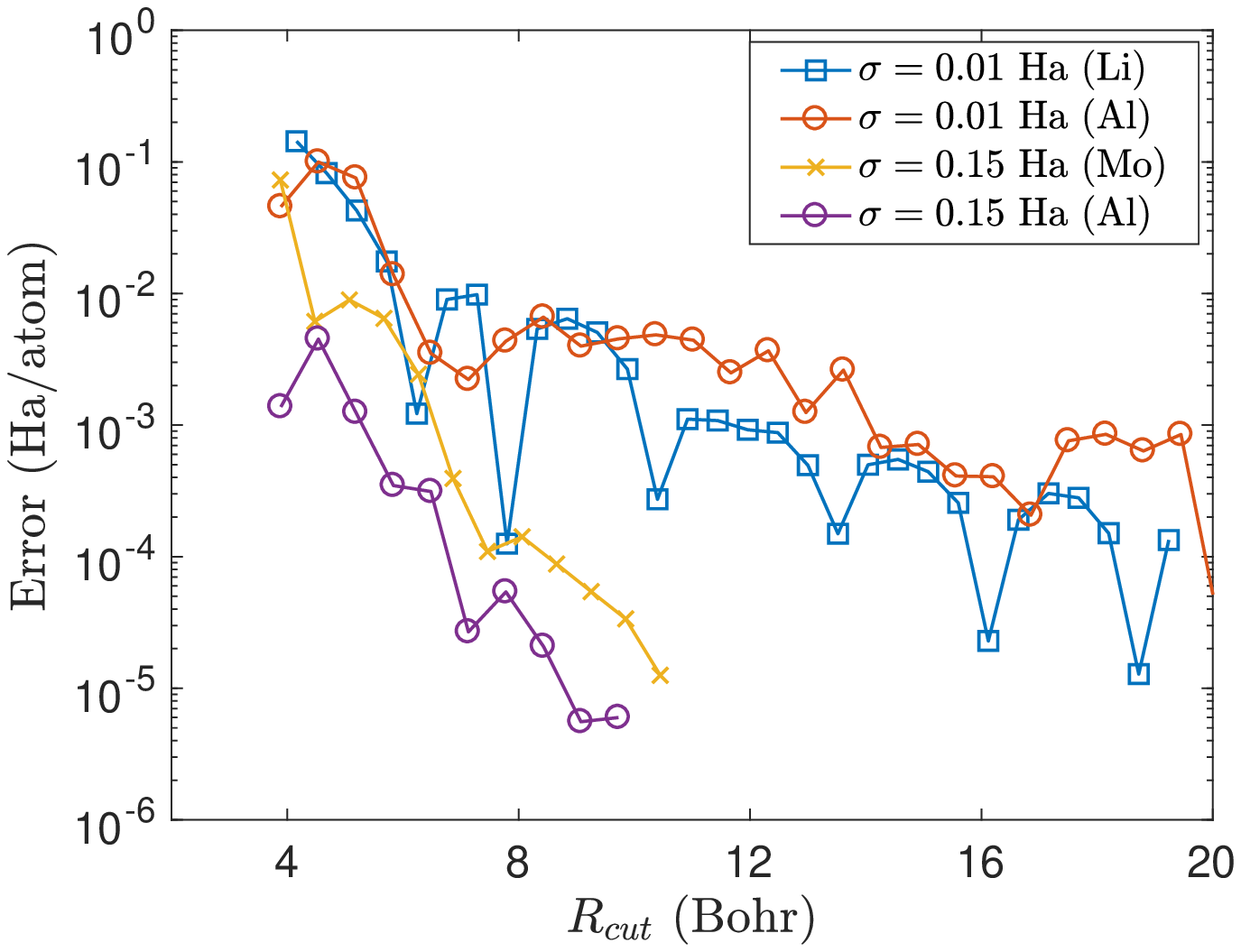}}
\subfloat[Atomic forces]{\label{Fig:Comparison_force_Al_Mo_Li}\includegraphics[keepaspectratio=true,width=0.49\textwidth]{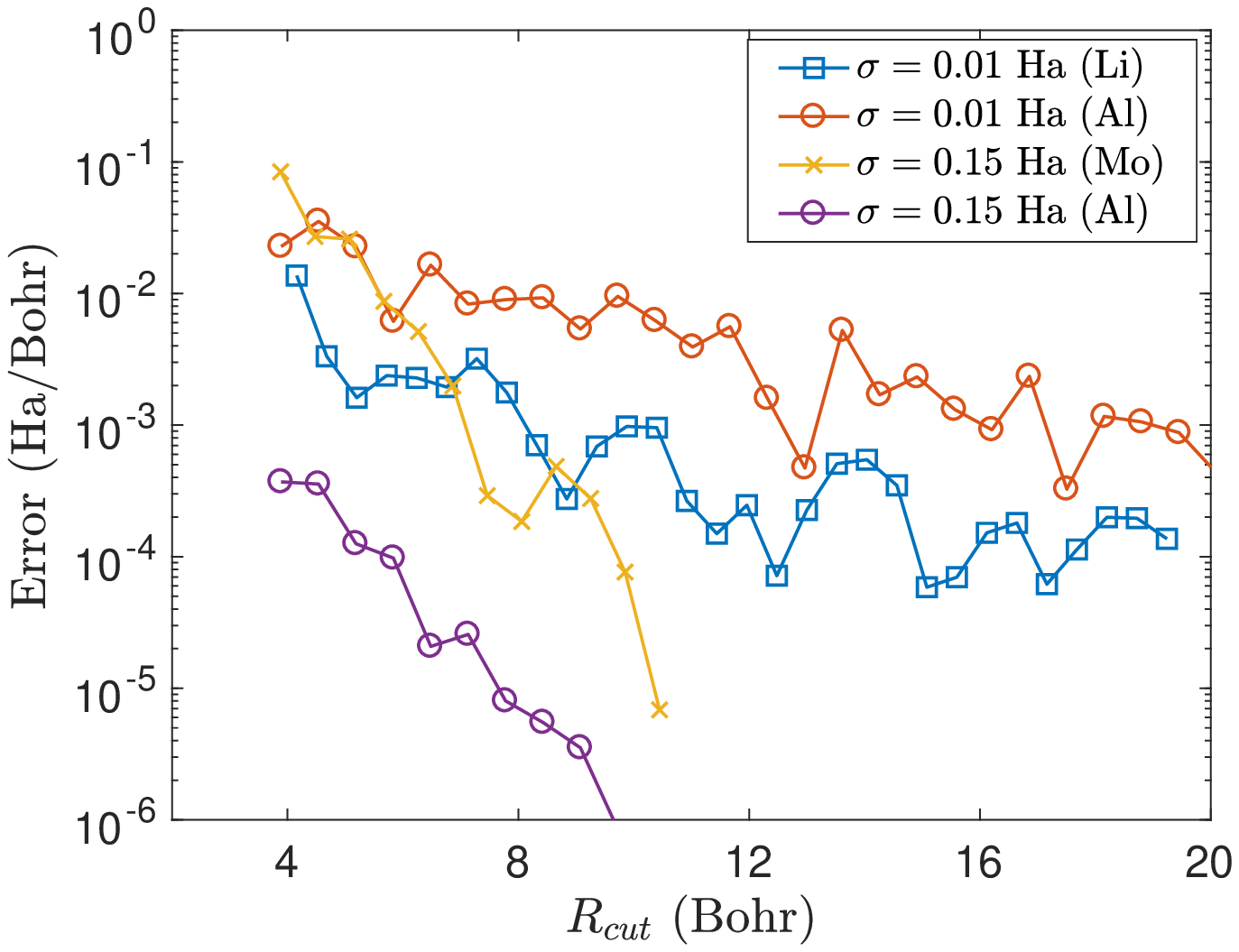}}
\caption{Convergence of energy and atomic forces with respect to truncation region size ($2 R_{cut}$) for the aluminum, lithium, and molybdenum systems. The results obtained by SPARC \cite{ghosh2016sparc1,ghosh2016sparc2} via diagonalization for the same mesh-size and finite-difference order are used as reference.}
\label{Fig:Comparison_Al_Li_Mo}
\end{figure}

In Fig. \ref{Fig:Comparison_Al_Li_Mo}, we present the error---defined with respect to the results obtained by SPARC \cite{ghosh2016sparc1,ghosh2016sparc2} via diagonalization at the same mesh-size and finite-difference order---in the energy and atomic forces as a function of the truncation region size ($2R_{cut}$) for the aforedescribed aluminum, lithium, and molybdenum systems. We observe that the convergence rate for lithium and molybdenum is very similar to that of aluminum. However, the associated prefactors are noticeably different, particularly in the case of the atomic forces. These results indicate that for a given smearing, the difference in the truncation region sizes required for achieving a desired accuracy in various metallic systems is likely to be consequence of the difference in the prefactor rather than the convergence rate. 

\subsection{Implications on O(N) DFT calculations}
The implication of the above results on practical $\mathcal{O}(N)$ DFT calculations for metallic systems merits further consideration. For aluminum, with generally used smearing of $\sigma = 0.001-0.01$ Ha, we have found that truncation regions of size $2R_{cut} = 48-64$ Bohr are required to achieve chemical accuracy of $0.001$ Ha/atom and $0.001$ Ha/Bohr in the energy and atomic forces, respectively. Consequently, $\mathcal{O}(N)$ scaling can be achieved in practical calculations only for domain sizes larger than $48-64$ Bohr, which translates to aluminum systems of $\mathcal{O}(1000)$ atoms. Indeed, the quantitative features of these results depend on the nature of the metallic system, as shown in the previous subsection. 

Finally, we note that the extremely rapid convergence of the energy and atomic forces with $R_{cut}$ for relatively large values of smearing suggests that non-orbital based $\mathcal{O}(N)$ methods like SQ are an extremely attractive choice for high-temperature simulations. For example, at $\sigma=0.15$ Ha smearing, truncation regions of size $2 R_{cut} \sim 8$ Bohr are sufficient to achieve an accuracy of $0.001$ Ha/atom and $0.001$ Ha/Bohr in the energy and atomic forces, respectively. This also suggests that orbital-free DFT \cite{WangBook2002,Suryanarayana2014524,ghosh2016higher}---$\mathcal{O}(N)$ theory in which the electronic kinetic energy is approximated with a functional of the electron density---becomes a better approximation to DFT at higher temperatures \cite{white2013orbital}, since the electronic kinetic energy can be expressed as the trace of the product of the density matrix with the Hamiltonian. 


\section{Concluding remarks} \label{Sec:Conclusions}
We have investigated the nearsightedness of electronic interactions in aluminum as a function of the smearing/electronic temperature in the context of $\mathcal{O}(N)$ Density Functional Theory (DFT) calculations. In particular, we have determined the variation of the error in the energy and atomic forces as a function of the truncation region size for smearing values of $0.001-0.15$ Ha. We have found that the convergence is exponential, with a rate whose growth increases with smearing, while remaining sub-linear. In particular, truncation regions of size $48-64$ Bohr are required to achieve chemical accuracy for typically used  smearing values of $0.001-0.01$ Ha. We have also found through comparison with lithium and molybdenum that the difference in truncation region sizes for various metallic systems is likely to be consequence of the difference in the prefactor rather than the convergence rate. This translates to very large prefactors for linear-scaling methods at moderate system sizes, and $\mathcal{O}(N)$ scaling in practical calculations only for systems larger than $\mathcal{O}(1000)$ atoms.


\section*{Acknowledgements}
This work was performed under the auspices of the U.S. Department of Energy by Lawrence Livermore National Laboratory under Contract DE-AC52-07NA27344. We gratefully acknowledge support from the Laboratory Directed Research and Development Program. We also acknowledge the valuable and insightful discussions with John E. Pask, and the help of Swarnava Ghosh and Phanisri P. Pratapa in performing some of the simulations relevant to this work. 


\bibliographystyle{ReferenceStyle}


\end{document}